\documentclass[a4paper,11pt]{article}
\usepackage[toc,page]{appendix}
\usepackage{graphicx}
\usepackage{colortbl}
\usepackage{amsmath}
\usepackage{subfigure}
\usepackage{mathtools}

\DeclarePairedDelimiterX\braket[2]{\langle}{\rangle}{#1 \delimsize\vert #2}
\usepackage[utf8]{inputenc}
\usepackage{float}
\usepackage[normalem]{ulem}
\usepackage{epstopdf}
\usepackage{amsfonts,amsmath,amssymb}
\usepackage{hyperref}

\usepackage{epsfig,multicol,bbm}
\usepackage{color}
\usepackage[dvipsnames]{xcolor}

\definecolor{darkblue}{rgb}{0.0, 0.0, 0.55}
\definecolor{grey}{rgb}{0.57, 0.64, 0.69}
\definecolor{lightbrown}{rgb}{0.71, 0.4, 0.11}

\newcommand{\be}{\begin{equation}}
\newcommand{\ee}{\end{equation}}

\def\ndelta{\delta\hspace{-0.50em}\slash\hspace{-0.05em} }

\newcommand\fverb{\setbox\pippobox=\hbox\bgroup\verb}
\newcommand\fverbit{\egroup\item[\fbox{\unhbox\pippobox}]}

\newbox\pippobox
\date{}
\begin{document}
\title{\bf More on Boundary Conditions for Warped AdS$_3$ in GMG}
\author{\textbf{S. N. Sajadi$^{a}$}\thanks{naseh.sajadi@gmail.com}, 
\textbf{Ali Hajilou$^{b}$}\thanks{hajilou@ipm.ir}
 \\\\
 \textit{{\small $^a$ Physics Department and Biruni Observatory, College of Sciences, Shiraz University, Shiraz 71454, Iran}}\\
\textit{{\small $^b$ School of Particles and Accelerators, Institute for Research in Fundamental Science (IPM),}}\\
\textit{{\small P.O. Box 19395-5746, Tehran, Iran}}
}
\maketitle
\begin{abstract}
In this paper, we study the  Aggarwal, Ciambelli, Detournay, and Somerhausen (ACDS) boundary conditions \cite{Aggarwal:2020igb} for Warped AdS$_3$ (WAdS$_3$) in the framework of General Massive Gravity (GMG) in the quadratic ensemble. We construct the phase space, the asymptotic structure, and the asymptotic symmetry algebra. We show that the global surface charges are finite, but not integrable, and also we find the conditions to make them integrable.
In addition, to confirm that the phase space has the same symmetries as that of a Warped Conformal Field Theories (WCFT), we compare the bulk entropy of Warped BTZ (WBTZ) black holes with the number of states belonging to a WCFT.
\end{abstract}

\maketitle
\section{Introduction}
One of the interesting achievements of string theory in the last two decades is the Anti de-Sitter/Conformal Field Theory (AdS/CFT) correspondence. This correspondence has opened a new approach to studying two different areas of physics, i.e. quantum field theory and gravity theory. After the introduction of the AdS/CFT correspondence in \cite{Maldacena:1997re,Witten:1998qj,Gubser:1998bc}, or more generally gauge/gravity duality, many different questions on the field theory side have been investigated utilizing the gravity side \cite{Sakai:2005yt}-\cite{Masoumi:2020cbj} (For more details see \cite{Casalderrey-Solana:2011dxg} and references therein). This duality proposes a correspondence between a quantum field theory in d-dimensional space-time and gravity theory in (d+1)-dimensional space-time. Fields, parameters, and quantities on the gauge theory side are translated to equivalent quantities on the gravity side. For instance, the vacuum state and thermal state on the field theory side correspond to the pure-AdS and black hole on the gravity theory, respectively. In addition, an extension of AdS/CFT correspondence to non-AdS geometries is Flat/Bondi-Metzner-Sachs invariant field theories (Flat/BMSFT) correspondence. According to this duality, asymptotically flat spacetimes in (d+1) dimensions are dual to d-dimensional BMSFTs \cite{Bagchi:2010zz}-\cite{Barnich:2012aw}.\\
As we know, the study of asymptotic symmetries in gravity theories is an old topic that has recently received attention. 
In the context of the AdS/CFT correspondence, the asymptotic symmetries of the gravity theory in the bulk spacetime correspond to the global symmetries of the dual quantum field theory in the boundary through the holographic dictionary. Therefore, with strong control of asymptotic symmetries, new holographic dualities can be investigated. The asymptotic symmetries are bulk residual transformations that preserve the boundary conditions but change the asymptotic field space (that is, they have non-vanishing surface charges). The asymptotic symmetry group is the group of residual gauge diffeomorphisms preserving the boundary conditions with associated non-vanishing charges. 
The boundary conditions determine the structure of the asymptotic symmetry group. Brown and Henneaux studied asymptotic symmetries of three-dimensional AdS space (AdS$_3$) and found that the symmetry algebra forms two copies of Virasoro algebra with a non-vanishing central charge ($c=3l/2G$) \cite{Brown:1986nw}. This implies that bulk theories with these boundary conditions are dual to CFTs with this central charge. Strominger and collaborators have extended these results to extremal Kerr black holes in what is known as the Kerr/CFT correspondence. Compere, Song, and Strominger (CSS) \cite{Compere:2013bya} have demonstrated a family of specific alternative boundary conditions in which the asymptotic symmetry algebra of a $3D$ theory turns out to consist of a semi-direct product of a Virasoro and $u(1)$ Kac-Moody algebras which are symmetries of the 2-dimensional WCFT’s (that is invariant under chiral scaling and translations but not rotations).  In \cite{Ciambelli:2020shy} and \cite{Setare:2021ugr}, Topologically Massive Gravity (TMG)  and General Minimal Massive Gravity (GMMG) with the CSS boundary conditions are studied. We refer interested reader to \cite{Adami:2020ugu}-\cite{Arefeva:2018hyo} for more details.

In \cite{Aggarwal:2020igb}, a new set of boundary
conditions in three-dimensional TMG has been introduced so that the dual field theory is a WCFT in the quadratic ensemble\footnote{Quadratic ensemble is similar to the canonical ensemble with different zero modes of their algebra.}. The boundary conditions of \cite{Aggarwal:2020igb} generalize those of \cite{Compere:2013bya} by accommodating more solutions. In this work, along the line of work \cite{Aggarwal:2020igb}, we introduce the ACDS boundary conditions and study its consequences like solutions space, asymptotic symmetries, and charge algebra in the framework of the GMG theory. In fact, we extended the domain of validity of these new boundary conditions to GMG theory.

The rest of this paper is organized as follows: In section \ref{sec2}, after the introduction of the GMG theory and boundary conditions, we impose the field equations to determine the solution space. In section \ref{secc3}, we compute the asymptotic Killing vectors preserving the boundary conditions and  the gauge and their corresponding surface charges in GMG. We find that the surface charges are not integrable, but by fixing a part of the solution space one can obtain the integrable charges.
In section \ref{secc4}, we compute the bulk thermodynamic entropy and compare it with the WCFT Cardy formula, showing that they match once the vacuum is correctly identified.
Finally, we provide some conclusions in Section \ref{secc5}.

\section{GMG under ACDS boundary conditions}\label{sec2}
The generalized massive gravity theory is realized by adding both the Chern-Simons (CS) deformation term and the higher derivative deformation term to pure Einstein gravity with a negative cosmological constant. This theory has two mass parameters and TMG and New Massive Gravity (NMG) are just two different limits of this generalized theory\cite{Sinha:2010ai}-\cite{Deser:1982vy}. The action for the generalized massive gravity theory can be written as \cite{Bergshoeff:2009aq},\cite{Liu:2009pha} 
\begin{equation}\label{action}
S_{GMG}=\dfrac{1}{8\pi G}\int d^{3}x\sqrt{-g}\left[s \mathcal{R}-2\lambda +\dfrac{1}{\mu}\mathcal{L}_{CS}+\dfrac{1}{\zeta^2}\mathcal{L}_{NMG}\right],
\end{equation}
where
\begin{equation}
\mathcal{L}_{CS}=\dfrac{1}{2}\epsilon^{\lambda \mu \nu}\left( \Gamma^{\rho}_{\lambda \sigma}\partial_{\mu}\Gamma^{\sigma}_{\rho \nu}+\dfrac{2}{3}\Gamma^{\rho}_{\lambda \sigma}\Gamma^{\sigma}_{\mu \tau}\Gamma^{\tau}_{\rho \nu}\right),\;\;\;\;\;\;\mathcal{L}_{NMG}=
\mathcal{R}^{\mu \nu}\mathcal{R}_{\mu \nu}-\dfrac{3}{8}\mathcal{R}^{2},
\end{equation}
and $\mu$ and $\zeta$ are the mass parameters of TMG and NMG, respectively. $\lambda$ is a cosmological parameter with the dimension of mass squared, and $s$ is a  conventional sign.
Varying the action with respect to the metric, one gets the following equations of motion
\begin{equation}\label{eq1}
\mathcal{E}_{\mu \nu}=\bar{s}\mathcal{G}_{\mu \nu}+\bar{\lambda} g_{\mu \nu}+\dfrac{1}{\mu}\mathcal{C}_{\mu \nu}+\dfrac{1}{2\zeta^{2}}\mathcal{K}_{\mu \nu},
\end{equation}
where $\mathcal{G}_{\mu\nu}$ is the Einstein tensor, $\mathcal{C}_{\mu \nu}$ the Cotton tensor, and $\mathcal{K}_{\mu\nu}$ is given by
\begin{equation}\label{eq2}
\mathcal{K}_{\mu \nu}=-\dfrac{1}{2}\nabla^{2}\mathcal{R} g_{\mu \nu}-\dfrac{1}{2}\nabla_{\mu}\nabla_{\nu} \mathcal{R}+2\nabla^{2}\mathcal{R}_{\mu \nu}+4 \mathcal{R}_{\mu a \nu b}\mathcal{R}^{a b}-
\dfrac{3}{2}\mathcal{R} \mathcal{R}_{\mu \nu}-\mathcal{R}_{\alpha \beta}\mathcal{R}^{\alpha \beta}g_{\mu \nu}+\dfrac{3}{8}\mathcal{R}^{2}g_{\mu \nu}.
\end{equation}
The parameters $\bar{s}$ and $\bar{\lambda}$ are parameters defined in terms of other parameters like $s$, $\mu$ and $\zeta$.
The Fefferman-Graham gauge in three spacetime dimensions in coordinates $ x^{\mu}=(r, x^{+},x^{-}) $ with three gauge-fixing conditions is
\begin{equation}
g_{rr}=\dfrac{L^2}{r^2},\;\;\;\;\; g_{r a}=0,
\end{equation}
where $x^{\pm}=t/L\pm \phi$.
 The line element takes the form
\begin{equation}\label{metric}
ds^2=\dfrac{L^2}{r^2}dr^2+\gamma_{a b}(r, x)dx^{a}dx^{b}.
\end{equation}
We consider the following fall-offs of the metric \cite{Aggarwal:2020igb}
\begin{equation}\label{boundcond}
 \gamma_{r r}=\dfrac{L^2}{r^2}+\mathcal{O}(r^{-4}),\;\;\;\gamma_{+ +}=\mathcal{O}(r^4),\;\;\;\gamma_{+ -}=\mathcal{O}(r^2),\;\;\;\;\gamma_{- -}=\mathcal{O}(1).
\end{equation}
Therefore, the field equations (\ref{eq1}) give us
\begin{align}
\gamma_{+ +}=&j_{++}r^{4}+h(x^{+})r^2+f_{++}(x^{+})+\dfrac{(1-A^{2})((A^2-1)h(x^{+})^{2}+4j_{++}f_{++}(x^{+}))h(x^{+})}{8j_{++}^{2}r^2(1+A^{2})}\nonumber\\
&+\dfrac{(A^2-1)^{2}(4j_{++}f_{++}(x^{+})+(A^2-1)h^{2}(x^{+}))^{2}}{64(1+A^{2})^{2}j_{++}^{3}r^{4}},\\
\gamma_{+ -}=&\varsigma_{+ -} r^2+\dfrac{(1-A^2)h(x^{+})\varsigma_{+ -}}{2j_{++}}+\dfrac{(1-A^{2})((A^2-1)h(x^{+})^{2}+4j_{++}f_{++}(x^{+}))\varsigma_{+-}}{8j_{++}^{2}(1+A^{2})r^{2}},\\
\gamma_{- -}=&\dfrac{(1-A^{2})\varsigma_{+ -}^{2}}{j_{++}},
\end{align}
with
\begin{align}
\bar{\lambda}=&\dfrac{1}{3087\mu^4L^4\zeta^{2}}[2L\zeta^{2}(3L^2\zeta^{4}-14\bar{s} L^2\mu^2\zeta^{2}-56\mu^2)\sqrt{84\mu^2+\zeta^{2}L^2(9\zeta^{2}-42\bar{s})}-
18L^{4}\zeta^{8}\nonumber\\
&+126\bar{s} \mu^2 \zeta^{6}L^{4}+(252-147\bar{s}^2\mu^2L^2)\mu^2L^2\zeta^{4}-4704\bar{s} \mu^4\zeta^{2}L^2-2352\mu^4].
\end{align}
The Ricci scalar of \eqref{metric} is given by
\begin{equation}
\mathcal{R}=\dfrac{2(1-4A^2)}{A^2L^2}=6\bar{\lambda},
\end{equation}
where\footnote{\it RootOf is a command used as a placeholder
for roots of equations in Maple \cite{rootof}. }
\begin{small}
\begin{equation}\label{eqaaa}
A={\it RootOf}\left[(16\mu+4\lambda{L}^{4}\mu{\zeta}
^{2}) {\_Z}^{4}+63\mu-16 L{\zeta}^{2}{\_Z}+
 (-80\mu +4\mu{L}^{2}{\zeta}^{2}\bar{s}) {\_Z
}^{2}+16{\zeta}^{2}L{\_Z}^{3},{\it label}={\_L3}
 \right].
\end{equation}
\end{small}
Therefore, it is negative as long as $\bar{\lambda}$ is. As it can be shown, the solution space is characterized by four quantities: two constants $j_{++}$ and $\varsigma_{+-}$ and two functions $h(x_{+})$ and $f_{++}(x_{+})$. 
By writing the WBTZ black holes (\ref{WBTZsol}) in the Fefferman-Graham gauge with the boundary conditions \eqref{boundcond}, one gets
\begin{align}
h=&~0~,\label{eqq14wbtz1}\\
j_{++}=&-\dfrac{119\mu^2 -6\zeta^4 L^2 +2\zeta^2 L \sqrt{84\mu^2
+\zeta^2 L(9\zeta^2 -42\bar{s}\mu^2)}+14\bar{s}\mu^2\zeta^2 L^2}{1176GL(LM-J)\mu^2},\\ \varsigma_{+-}=&\dfrac{1}{43218\mu^4}[4L\zeta^2(-119\mu^2+6\zeta^4L^2-14\bar{s} \mu^2\zeta^2L^2)\sqrt{84\mu^2+\zeta^2 L^2(9\zeta^2-42\mu^2\bar{s})}+29057 \nonumber\\
&\mu^4 -4\zeta^2L^2(18L^2\zeta^6-84\bar{s} L^2\mu^2\zeta^4-273\zeta^2\mu^2+49\bar{s}^2L^2\zeta^2\mu^4+833\bar{s} \mu^4)],\\
\gamma_{--}=&\dfrac{GL(LM-J)}{21609\mu^4}[4L\zeta^2(-119\mu^2+6\zeta^4L^2-14\bar{s} \mu^2\zeta^2L^2)\sqrt{84\mu^2+\zeta^2 L^2(9\zeta^2-42\mu^2\bar{s})}+\nonumber\\
&29057\mu^4-4\zeta^2L^2(18L^2\zeta^6-84\bar{s} L^2\mu^2\zeta^4-273\zeta^2\mu^2+49\bar{s}^2L^2\zeta^2\mu^4+833\bar{s} \mu^4)],\\
f_{++}=&\dfrac{GL (ML+J)}{43218\mu^4}[4L\zeta^2(-119\mu^2+6\zeta^4L^2-14\bar{s} \mu^2\zeta^2L^2)\sqrt{84\mu^2+\zeta^2 L^2(9\zeta^2-42\mu^2\bar{s})}+\nonumber\\
&72275\mu^4-4\zeta^2L^2(18L^2\zeta^6-84\bar{s} L^2\mu^2\zeta^4-273\zeta^2\mu^2+49\bar{s}^2L^2\zeta^2\mu^4+833\bar{s} \mu^4)].\label{eqq14wbtz2}
\end{align}
In the case of $A=1$ and arbitrary $j_{++}$, the metric becomes
\begin{equation}
ds^2=\dfrac{L^2}{r^2}dr^2+(j_{++}r^{4}+h(x^{+})r^2+f_{++}(x^{+}))dx^{+2}+\varsigma_{+-}r^2dx^{+}dx^{-},
\end{equation}
and 
\begin{equation}
\bar{\lambda}=-\dfrac{12L\zeta^2-35\mu}{4\mu L^4\zeta^2},\;\;\;\; \bar{s}=\dfrac{6\zeta^2 L-17\mu}{2\mu L^2\zeta^2}.
\end{equation}
This metric is not a solution for the Einstein equation, because the Cotton tensor and NMG part have non-vanishing components ($\mathcal{C}^{+}_{-}=\frac{12~r^2 j_{++}}{\varsigma_{+-}L^{3}},\mathcal{K}^{+}_{-}=-\frac{136~r^2j_{++}}{\varsigma_{+-}L^4}$).
In the case of $A=1$, $\mu=\frac{6~\zeta^2 L}{17+2\sigma \zeta^2L^2}$ and $j_{++}\to 0$ but keeping the ratio $\Delta =\frac{A^2-1}{j_{++}}$ constant, the line element becomes
\begin{align}
ds^{2}=&\dfrac{L^2}{r^2}dr^2+\left[h(x^{+})r^2+f_{++}(x^{+})+\dfrac{h(x^{+})\Delta [4f_{++}(x^{+})+\Delta h^{2}(x^{+})]}{16r^2}\right]dx^{+2}+\Delta \varsigma_{+-}^{2}dx^{-2}\nonumber\\
&+\left[\varsigma_{+-}r^2+\dfrac{h(x^{+})\varsigma_{+-}\Delta}{2}+\dfrac{\varsigma_{+-}\Delta(4f_{++}(x^{+})+h^{2}(x^{+})\Delta)}{16r^2}\right]dx^{+}dx^{-}.
\end{align}
This metric is the CSS metric \cite{Compere:2013bya}. 
\section{Symmetries and charges}\label{secc3}
The residual gauge diffeomorphisms are generated by the vector $\xi$ satisfying
\begin{equation}
\mathcal{\L}_{\xi}g_{rr}=0,\;\;\;\;\; \mathcal{\L}_{\xi}g_{ra}=0,
\end{equation}
where $\mathcal{\L}$ denotes the Lie derivative.
The solutions to these equations are
\begin{equation}\label{eqqkil}
\xi =\xi^{\mu}\partial_{\mu}=\xi^{r}\partial_{r}+\xi^{+}\partial_{+}+\xi^{-}\partial_{-},
\end{equation}
with
\begin{equation}
\xi^{r}=r\eta(x^{+}),
\end{equation}

\begin{align}
\xi^{+}&=\epsilon +\dfrac{2L^2j_{++}\eta^{\prime}(A^{4}-1)}{A^{2}((A^{2}-1)^{2}h^{2}+4j_{++}f_{++}(A^{2}-1)+8r^4j_{++}^{2}(A^{2}+1))},\nonumber\\
\xi^{-}&=\sigma - \dfrac{j_{++}L^2\eta^{\prime}(A^2+1)[(A^2-1)h-4j_{++}r^2]}{A^{2}\varsigma_{+-}[(A^2-1)^{2}h^2+4j_{++}f_{++}(A^2-1)+Aj_{++}^{2}r^4(A^2+1)]}.
\end{align}
In these expressions, $\sigma(x^{+})$ and $\epsilon(x^{+})$ are field-independent arbitrary functions.
Varying the metric (\ref{metric}) along $\xi$, we find the variation of solution space as follows
\begin{equation}
\mathcal{\L}_{\xi}g_{\mu \nu}dx^{\mu}dx^{\nu}=\dfrac{L^2}{r^2}dr^2+\delta_{\xi}\gamma_{a b}(r,x)dx^{a}dx^{b},
\end{equation}
with 
$\delta_{\xi}\gamma_{++}=\L_{\xi}\gamma_{++}$ , 
then we have
\begin{align}
\delta_{\xi}j_{++}=& 2j_{++}(\epsilon^{\prime}+2\eta)\label{eqdelta30}\\
\delta_{\xi}h=& 2h\eta +2\varsigma_{+-}\sigma^{\prime}+\epsilon h^{\prime}+2h\epsilon^{\prime}\\
\delta_{\xi}f_{++}=&\epsilon f_{++}^{\prime}+2f_{++}\epsilon^{\prime}+\dfrac{L^2\eta^{\prime\prime}(A^2+1)}{2A^{2}}+\dfrac{h\sigma^{\prime}\varsigma_{+-}(1-A^2)}{j_{++}},
\end{align}
and 
\begin{align}\label{eqq20}
\delta_{\xi}\gamma_{+-}=\L_{\xi}\gamma_{+-}
\;\;\;\to\;\;\;\delta_{\xi}\varsigma_{+-}=\varsigma_{+-}(2\eta+\epsilon^{\prime}).
\end{align}
By requiring $j_{++}$ to be constant, from \eqref{eqdelta30} we get
\begin{equation}\label{eqq34}
\eta=-\dfrac{1}{2}\epsilon^{\prime}+\eta_{0}.
\end{equation}
Therefore, the transformation of $j_{++}$ becomes
\begin{equation}
\delta_{\xi}j_{++}=4j_{++}\eta_{0}.
\end{equation}
If we assume $\eta_{0}=0$, then we obtain 
\begin{equation}
\delta_{\xi}j_{++}=\delta_{\xi}\varsigma_{+-}=0.
\end{equation}
This means that $j_{++}$ and $\varsigma_{+-}$ are fixed along the residual orbits. Finally,
we find the full residual variation of solution space as
\begin{align}
\delta_{\xi}j_{++}&=0,\\
\delta_{\xi}h&=(h\epsilon)^{\prime}+2\varsigma_{+-}\sigma^{\prime},\\
\delta_{\xi}f_{++}&=\epsilon f_{++}^{\prime}+2f_{++}\epsilon^{\prime}-\dfrac{L^2\epsilon^{\prime\prime\prime}(A^2+1)}{4A^{2}}+\dfrac{h\sigma^{\prime}\varsigma_{+-}(1-A^2)}{j_{++}}.
\end{align}
The general symmetry generators, using \eqref{eqq34}, are as follows
\begin{equation}
\xi^{r}=-\dfrac{1}{2}r\epsilon^{\prime},\nonumber
\end{equation}
\begin{small}
\begin{align}\label{eqqxi}
\xi^{+}&=\epsilon -\dfrac{L^2j_{++}\epsilon^{\prime\prime}(A^{4}-1)}{A^{2}((A^{2}-1)^{2}h^{2}+4j_{++}f_{++}(A^{2}-1)+8r^4j_{++}^{2}(A^{2}+1))},\nonumber\\
\xi^{-}&=\sigma + \dfrac{j_{++}L^2\epsilon^{\prime\prime}(A^2+1)[(A^2-1)h-4j_{++}r^2]}{2A^{2}\varsigma_{+-}[(A^2-1)^{2}h^2+4j_{++}f_{++}(A^2-1)+Aj_{++}^{2}r^4(A^2+1)]},
\end{align}
\end{small}
where $A$ is defined in \eqref{eqaaa}.
The  residual symmetries \eqref{eqqxi} depend on two arbitrary chiral functions $\epsilon(x^{+})$ (generating the usual Witt algebra) and $\sigma(x^{+})$ (generating an abelian algebra). Therefore, the total asymptotic symmetry algebra is a direct sum of a Witt and a $u(1)$ algebra.
\subsection{Charges and Algebra}
The surface charges are computed using \cite{Nam:2010ub}, \cite{Setare:2022vme},\cite{Bouchareb:2007yx} as follows
\begin{align}\label{eq10}
\ndelta Q^{a}(\bar{\xi})=\int_{\Sigma}dS_{i}F^{a i}(g,h),
\end{align}
with
\begin{small}
\begin{align}
F_{E}^{a i}(\bar{\xi})=&\bar{\xi}_{b}\bar{\nabla}^{a}h^{i b}-\bar{\xi}_{b}\bar{\nabla}^{i}h^{a b}+\bar{\xi}^{a}\bar{\nabla}^{i}h-\bar{\xi}^{i}\bar{\nabla}^{a}h+h^{a b}\bar{\nabla}^{i}\bar{\xi}_{b}-h^{i b}\bar{\nabla}^{a}\bar{\xi}_{b}
 +\bar{\xi}^{i}\bar{\nabla}_{b}h^{a b}
 -\bar{\xi}^{a}\bar{\nabla}_{b}h^{i b}+h\bar{\nabla}^{a}\bar{\xi}^{i},\\ 
F_{C}^{a i}(\bar{\xi})=&F_{E}^{a i}(\eta)+\dfrac{1}{\sqrt{\bar{g}}}\bar{\xi}_{\lambda}\left(\epsilon^{a i \rho}\delta G^{ \lambda}_{\rho}-\dfrac{1}{2}\epsilon^{a i \lambda}\delta G\right)+\dfrac{1}{2\sqrt{{g}}}\epsilon^{a i \rho}\left[\bar{\xi}_{\rho}h^{\lambda}_{\sigma}{G}^{\sigma}_{\lambda}+\dfrac{1}{2}h\left(\bar{\xi}_{\sigma}{G}^{\sigma}_{\rho}+\dfrac{1}{2}\bar{\xi}_{\rho}{R}\right)\right],\\
F^{a b}_{R^{2}}(\bar{\xi})=&2RF^{a b}_{E}(\bar{\xi})+4\bar{\xi}^{[a}\nabla^{b]}\delta R+2\delta R\nabla^{[a}\bar{\xi}^{b]}-2\bar{\xi}^{[a}h^{b]\alpha}\nabla_{\alpha}R,\\
F^{a b}_{R_{2}}(\bar{\xi})=&\nabla^{2}F^{a b}_{E}+\dfrac{1}{2}F^{a b}_{R^{2}}-2F_{E}^{\alpha[a}R^{b]}_{\alpha}-2\nabla^{\alpha}\bar{\xi}^{\beta}\nabla_{\alpha}\nabla^{[a}h^{b]}_{\beta}-
4\bar{\xi}^{\alpha}R_{\alpha \beta}\nabla^{[a}h^{b]\beta}-Rh_{\alpha}^{[a}\nabla^{b]}\bar{\xi}^{\alpha}+
2\bar{\xi}^{[a}R^{b]}_{\alpha}\nabla_{\beta}\nonumber\\
&h^{\alpha \beta}+2\bar{\xi}_{\alpha}R^{\alpha[a}\nabla_{\beta}h^{b]\beta}+2\bar{\xi}^{\alpha}h^{\beta[a}\nabla_{\beta}R^{b]}_{\alpha}
+2h^{\alpha \beta}\bar{\xi}^{[a}\nabla_{\alpha}R^{b]}_{\beta}-
(\delta R+2~R^{\alpha \beta}h_{\alpha \beta})\nabla^{[a}\bar{\xi}^{b]}
-3~\bar{\xi}^{\alpha}\nonumber\\
&R^{[a}_{\alpha}\nabla^{b]}h-\bar{\xi}^{[a}R^{b]\alpha}
\nabla_{\alpha}h,\label{eqcharge42}
\end{align}
\end{small}%
where $\delta \mathcal{R}=-\mathcal{R}^{\alpha \beta}h_{\alpha \beta}+\nabla^{\alpha}\nabla^{\beta}h_{\alpha \beta}-\nabla^{2}h$, $\eta^{\nu}=\epsilon^{\nu \rho \sigma}\bar{\nabla}_{\rho}\bar{\xi}_{\sigma}$ and $h=\delta g_{\mu \nu}(\delta \alpha,\alpha)=\partial g_{\mu \nu}/\partial \alpha \delta \alpha$.
For the $u(1)$ sector and the Killing vector $\underline{\sigma}=\sigma(x^{+})\partial_{-}$, the surface charge becomes
\begin{small}
\begin{align}
\ndelta Q_{\underline{\sigma}}=&\dfrac{A^2-1}{8A^5\mu L^3\zeta^2 j_{++}^{2}}\int_{0}^{2\pi} d\phi \sigma(x^{+})[2(-8\mu A^4+4\zeta^2 L A^3+2\mu A^2(\bar{s} L^2 \zeta^2+34)-2\zeta^2 LA-63\mu)\varsigma_{+-}\nonumber\\
&j_{++}\delta h-j_{++}\delta\varsigma_{+-}((4(-14-\bar{s} L^2\zeta^2)\mu A^4+14LA^3\zeta^2 +2\mu A^2(4\bar{s} L^2 \zeta^2+151)-8\zeta^2 L A-63\mu) \nonumber\\
&h+(2\mu A^4(10+\bar{s} L^2\zeta^2)-3\zeta^2 LA^3-(\bar{s} L^2\zeta^2+83)\mu A^2+2\zeta^2 LA+63\mu)\varsigma_{+-})+(A^2-1)\varsigma_{+-}\nonumber\\
&\delta j_{++}(\varsigma_{+-}-h)(\mu A^2(10+\bar{s} L^2\zeta^2)-LA\zeta^2-63) ].
\end{align}
\end{small}
The Virasoro charges are
\begin{small}
\begin{align}
{\ndelta Q_{\underline{\epsilon}}}=&\dfrac{1}{16\mu L^3 \zeta^2 A^{5}(A^2+1)j_{++}^{2}\varsigma_{+-}}\int_{0}^{2\pi}d\phi (L^2 j_{++}(A^2+1)\epsilon^{\prime\prime}((A^2-1)(8LA^{3}\zeta^2-2\mu A^2(-14+\bar{s} L^2\zeta^2\nonumber\\
&-9AL^2\zeta^2+21\mu))\varsigma_{+-}\delta j_{++}-Aj_{++}\delta \varsigma_{+-}(10A^4L\zeta^2+A^{3}(40\mu+4\mu\bar{s} L^2\zeta^2)-25A^2L\zeta^{2}-42\mu A+\nonumber\\
&11L\zeta^2))-A^{2}\epsilon(A(A^2-1)\varsigma_{+-}j_{++}\delta h(-12A^4 L\zeta^2 -32A^3\mu(4+\bar{s} L^2\zeta^2)+10A^2L\zeta^2+L\zeta^2+144\mu \nonumber\\
&A)+\varsigma_{+-}j_{++}^2\delta f_{++}(24LA^5\zeta^2- 24\mu A^4(4+2\bar{s} L^2\zeta^2)+44A^3L\zeta^2)+8\mu A^2(35+2\bar{s} L^2\zeta^2)-20AL\zeta^2 \nonumber\\
&-168 \mu)-(A^2-1)\varsigma_{+-}\delta j_{++}(-Ah^2(12A^4 L\zeta^2+16\mu A^3(4+\bar{s} L^2\zeta^2- 10 A^2L\zeta^2 -72\mu A-2L\zeta^2)\nonumber\\
&+(A^2+1)\varsigma_{+-}h(16A^3 L\zeta^2 - 4AL\zeta^2 -84 \mu + 8 \mu A^{2}(10+\bar{s} L^2\zeta^2)-20(2A^2+1)ALf_{++}\zeta^2 j_{++})-\nonumber\\
&j_{++}\delta \varsigma_{+-}(12A^3(A^2-1)^2L\zeta^2 h^2-(A^4-1)\varsigma_{+-}h(16\mu A^2(10+\bar{s} L^2\zeta^2)+16AL\zeta^2-168\mu)+8AL \nonumber\\ 
&\zeta^2 f_{++} j_{++}(10A^4-5A^2-3)))))) ,
\end{align}
\end{small}%
where we have used $\underline{\epsilon}=\xi$, with $\xi^{r}$,  $\xi^{+}$ and  $\xi^{-}$ provided in \eqref{eqqxi} and $\sigma=0$. To obtain the above surface charges we evaluated equations \eqref{eq10}-\eqref{eqcharge42} first at ($r, x^{+}$) fixed, and second at ($r, x^{-}$) fixed,
 then added them together, and finally sent $r\to \infty$. These charges are finite but are not integrable. Non-integrability of charges implies that the finite charge expressions rely on the particular path that one chooses to integrate on the solution space, which is a common feature of a dissipative system. If $\delta j_{++}=\delta \varsigma_{+-}=0$ the charges become integrable. Also, one can find a combination of vectors such that these charges become integrable even when $\delta j_{++}\neq 0,\;\delta \varsigma_{+-}\neq0$.\\
Utilizing the integrable charges, the charge algebra is obtained. Therefore, in the case $\delta j_{++}=\delta \varsigma_{+-}=0$, the charges read as 
\begin{small}
\begin{align}
{\ndelta Q_{\underline{\sigma}}}=\dfrac{(A^2-1)\left(-8\mu A^4+4\zeta^2 L A^3+2\mu A^2(\bar{s} L^2 \zeta^2+34)-2\zeta^2 LA-{63\mu}\right)\varsigma_{+-}}{4A^5\mu L^3\zeta^2 j_{++}}\int_{0}^{2\pi} d\phi \sigma(x^{+})\delta h,
\end{align}
\end{small}
and
\begin{align}
\ndelta Q_{\underline{\epsilon}}=&-\dfrac{1}{8j_{++}\mu A^3 L^3\zeta^2 (A^2+1)}\int_{0}^{2\pi}d\phi\; \epsilon[A(A^2-1)\delta h(h(-6A^4 L\zeta^2-16A^3\mu(4+\bar{s} L^2\zeta^2)+ \nonumber\\
&5A^2L\zeta^2+72\mu A+L\zeta^2)+(16A^2-14)\varsigma_{+-}L\zeta^2 (A^2+1))+j_{++}\delta f_{++}(12L\zeta^2 A^5-24\mu \nonumber\\
&A^4(2+\bar{s} L^2\zeta^2)+22LA^3\zeta^2+4\mu A^2(35+2\bar{s} \zeta^2 L^2)-10AL\zeta^2-84\mu)] .
\end{align}
These charges can now be integrated. Integrating them, one obtains
\begin{small}
\begin{align}
 Q_{\underline{\sigma}}=\dfrac{(A^2-1)\left(-8\mu A^4+4\zeta^2 L A^3+2\mu A^2(\bar{s} L^2 \zeta^2+34)-2\zeta^2 LA-{63\mu}\right)\varsigma_{+-}}{4A^5\mu L^3\zeta^2 j_{++}}\int_{0}^{2\pi} d\phi \sigma(x^{+})(h+h_{0}) ,
\end{align}
\end{small}
and
\begin{align}
{ Q_{\underline{\epsilon}}}=&-\dfrac{1}{16j_{++}\mu A^3 L^3\zeta^2 (A^2+1)}\int_{0}^{2\pi}d\phi\; \epsilon[A(A^2-1)(h^2(-6A^4 L\zeta^2-16A^3\mu(4+\bar{s} L^2\zeta^2)+\nonumber\\
&5A^2L\zeta^2 +72 \mu A+L\zeta^2)+2(16A^2-14)\varsigma_{+-}hL\zeta^2 (A^2+1))+2j_{++}f_{++}(12L\zeta^2 A^5-24 \nonumber\\
&\mu A^4(2+\bar{s} L^2\zeta^2)+22LA^3\zeta^2+4\mu A^2(35+2\bar{s} \zeta^2 L^2)-10AL\zeta^2-84\mu)] .
\end{align}
For the $u(1)$ sector, the charge algebra is computed as
\begin{equation}
\delta_{\sigma_{2}} Q_{\sigma_{1}}[g]=Q_{[\sigma_{1},\sigma_{2}]}+K_{\sigma_{1},\sigma_{2}}.
\end{equation}
Since $Q_{[\sigma_{1},\sigma_{2}]}=0$, the central extension for the $u(1)$ sector is
\begin{small}
\begin{equation}
K_{\sigma_{1},\sigma_{2}}=\dfrac{(A^2-1)\left(-8\mu A^4+4\zeta^2 L A^3+2\mu A^2(\bar{s} L^2 \zeta^2+34)-2\zeta^2 LA-{63\mu}\right)\varsigma_{+-}^{2}}{2A^5\mu L^3\zeta^2 j_{++}}\int_{0}^{2\pi} d\phi \sigma_{1}\sigma^{\prime}_{2}.
\end{equation}
\end{small}
Using the mode decomposition $\sigma_{1}=e^{imx^{+}}$, $\sigma_{2}=e^{inx^{+}}$, and calling $ Q_{\underline{\sigma}^{1}}=P_{m} $, $ Q_{\underline{\sigma}^{2}}=P_{n} $, it is easy to obtain
\begin{small}
\begin{equation}
i\left\lbrace P_{m},P_{n} \right\rbrace =m\dfrac{k}{2}\delta_{m+n,0},
\end{equation}
\end{small}%
where
\begin{small}
\begin{equation}
 k=\dfrac{2\pi(A^2-1)\left(-8\mu A^4+4\zeta^2 L A^3+2\mu A^2(\bar{s} L^2 \zeta^2+34)-2\zeta^2 LA-{63\mu}\right)\varsigma_{+-}^{2}}{A^5\mu L^3\zeta^2 j_{++}}.
\end{equation}
\end{small}%
This is a centrally extended $u(1)$ algebra with central extension $k$ called the Kac-Moody level. 
For the Virasoro sector we have:
\begin{align}
\left\lbrace Q_{\underline{\epsilon}^{1}},Q_{\underline{\epsilon}^{2}} \right\rbrace =&\delta_{\epsilon_{2}} Q_{\epsilon_{1}}[g] \nonumber\\
=&\dfrac{1}{16j_{++}\mu A^3 L^3\zeta^2 (A^2+1)}\int_{0}^{2\pi}d\phi \left(\epsilon_{1}\epsilon_{2}^{\prime}-\epsilon_{2}\epsilon_{1}^{\prime}\right)[A(A^2-1)(h^2(-6A^4 L\zeta^2 -16 \nonumber\\
&A^3 \mu(4+\bar{s} L^2\zeta^2)+5A^2L\zeta^2+72\mu A+L\zeta^2)+2(16A^2-14)\varsigma_{+-}hL\zeta^2 (A^2+1)]\nonumber\\
&- [12L \zeta^2 A^5-24\mu A^4(2+\bar{s} L^2\zeta^2)+22LA^3\zeta^2 +4\mu A^2(35+2\bar{s} \zeta^2 L^2)-10AL \nonumber\\
&\zeta^2 -84\mu ] \left( \dfrac{1}{32\mu L A^3 \zeta^{2}}\right)
\int_{0}^{2\pi}d\phi \epsilon_{1}\epsilon_{2}^{\prime\prime\prime}.
\end{align}
Using the mode decomposition representation $\epsilon_{1}=e^{imx^{+}}$, $\epsilon_{2}=e^{inx^{+}}$, and calling $Q_{\underline{\epsilon}_{1}}=L_{m}$, $Q_{\underline{\epsilon}_{2}}=L_{n}$, one obtains
\begin{align}
i\left\lbrace L_{m}, L_{n} \right\rbrace =(m-n)L_{m+n}+\dfrac{c}{12}m^{3}\delta_{m+n,0},
\end{align}%
where
\begin{small}
\begin{equation}
 c=\dfrac{3\pi(12L\zeta^2 A^5-24\mu A^4(2+\bar{s} L^2\zeta^2)+22LA^3\zeta^2+4\mu A^2(35+2\bar{s} \zeta^2 L^2)-10AL\zeta^2-84\mu)}{4\mu L A^3 \zeta^{2}}.
\end{equation}
\end{small}%
In summary, the algebra is
\begin{align}
i\left\lbrace L_{m}, L_{n} \right\rbrace &=(m-n)L_{m+n}+\dfrac{c}{12}m^{3}\delta_{m+n,0},\label{eqqalgebra1}\\
i\left\lbrace L_{m}, P_{n} \right\rbrace &=-nP_{n+m}\\
i\left\lbrace P_{m},P_{n} \right\rbrace &=m\dfrac{k}{2}\delta_{m+n,0}\label{eqqalgebra2}
\end{align}
with central extensions
\begin{align}\label{centralex}
c&=\dfrac{3\pi(12L\zeta^2 A^5-24\mu A^4(2+\bar{s} L^2\zeta^2)+22LA^3\zeta^2+4\mu A^2(35+2\bar{s} \zeta^2 L^2)-10AL\zeta^2-84\mu)}{4\mu L A^3 \zeta^{2}},\nonumber\\k&=\dfrac{2\pi(A^2-1)\left(-8\mu A^4+4\zeta^2 L A^3+2\mu A^2(\bar{s} L^2 \zeta^2+34)-2\zeta^2 LA-{63\mu}\right)\varsigma_{+-}^{2}}{A^5\mu L^3\zeta^2 j_{++}}.
\end{align}
Therefore, from \eqref{eqqalgebra1}-\eqref{eqqalgebra2} with the associated central charges \eqref{centralex}, the bulk solution space has a symmetry algebra identified with that of a WCFT in the quadratic ensemble.

In the limit $\zeta\to \infty$, and $\mu>0,L>0$ one obtains \cite{Aggarwal:2020igb}
\begin{equation}
c=\dfrac{\mu^2 L^2\bar{s}^2 +9}{3\mu},\;\;\;\;\;k=-\dfrac{\varsigma_{+-}^2(\mu^2L^2\bar{s}^2-9)}{\mu L^2j_{++}},
\end{equation}
while in the case of $\mu\to \infty$, we have \cite{NMGpaper}
\begin{equation}
c=\dfrac{16\pi}{7}\sqrt{\dfrac{2}{21}}\dfrac{(\zeta^2L^2+2)^{\frac{3}{2}}}{\zeta^2L},\;\;\;\;\; k=-\dfrac{8\pi\sqrt{42}(2\zeta^2L^2-17)\varsigma_{+-}^{2}}{21\zeta^2L^3\sqrt{\zeta^2L^2+2}j_{++}}.
\end{equation}
This algebra is one of the centrally extended group
\begin{equation}
Vir\otimes U(1).
\end{equation}
Now, we study the null warped limit in the case $A=1$. In this case, the $u(1)$ level and charges vanish identically ($k=0$), we are left with a Virasoro symmetry algebra with central extension

\begin{equation}
c=\dfrac{18\pi(6\mu -L\zeta^2)}{\mu L\zeta^2},\;\;\;\; A \to 1.
\end{equation}
As we know, the CSS limit can be achieved setting $A=1$ and $j_{++}=0$ while keeping $\Delta=\frac{A^2-1}{j_{++}}$ constant. The charges read
{
\begin{align}
{ Q_{\underline{\sigma}}}&=\dfrac{\Delta \varsigma_{+-}(2\zeta^2 L-5\mu)}{\mu L^3 \zeta^2}\int_{0}^{2\pi} d\phi \sigma(x^{+})(h+h_{0}),\\
{ Q_{\underline{\epsilon}}}&=-\dfrac{1}{4\mu L^3 \zeta^2}\int_{0}^{2\pi}d\phi\epsilon(x^{+}) \left[6f_{++}(6\mu-\zeta^2 L)+\Delta(6h^2(3\mu-\zeta^2 L)+\varsigma_{+-}L\zeta^2 h)\right],
\end{align}}
while the central extensions become
\begin{equation}
c=\dfrac{18\pi(6\mu-\zeta^2 L)}{\mu L\zeta^2},\;\;\;\;k=\dfrac{8\pi \Delta \varsigma_{+-}^{2}(2\zeta^2 L-5\mu)}{\mu L^3\zeta^2}.
\end{equation}
This limit coincides with our results in \cite{Setare:2021ugr}.
We now turn our attention to the solution space of WBTZ black holes in \eqref{eqq14wbtz1}-\eqref{eqq14wbtz2}. 
Therefore, their charges in the quadratic ensemble take the form
\begin{align}
P_{m}=&\dfrac{2\pi h_{0}(A^2-1)(LM-J)(2H^2-1)}{A^5\mu L^2\zeta^2 H^2}
[-8\mu A^4+4\zeta^2 L A^3+2\mu A^2(\bar{s} L^2 \zeta^2+34)-2\zeta^2 LA \nonumber\\
&-{63 \mu} ] \delta_{m,0} ~ ,\\
L_{m}=&\dfrac{\pi G(J+ML)(H^2-1)}{2\mu A^3 L^2\zeta^2 (A^2+1)}[12L\zeta^2 A^5-24\mu A^4(2+\bar{s} L^2\zeta^2)+22LA^3\zeta^2+4\mu A^2(35+2\bar{s} \zeta^2\nonumber\\
& L^2)-10AL\zeta^2-84\mu]\delta_{m,0}~,
\end{align}
where $M$ and $J$ are Einstein charges. The GMG mass and angular momentum of these solutions are defined as
\begin{equation}
\mathcal{M}=Q_{\partial_{t}}=\dfrac{1}{L}(Q_{\partial_{+}}+Q_{\partial_{-}}),\;\;\;\; \mathcal{J}
=Q_{\partial_{\phi}}=Q_{\partial_{+}}-Q_{\partial_{-}},
\end{equation}
and we also have
\begin{equation}
Q_{\partial_{-}}=P_{0},\;\;\;\;Q_{\partial_{+}}=L_{0}.
\end{equation}
Then, the relation between the GMG mass, angular momentum, and the zero modes of the charges can be obtained as follows
\begin{align}\label{eqmassang}
\mathcal{M}=&\dfrac{1}{L}(P_{0}+L_{0})\nonumber\\
=&\dfrac{\pi}{2\mu L^3\zeta^2 A^{5}H^2(A^2+1)}[4h_{0}(A^4-1)(2H^2-1)(LM-J)(-8\mu A^4+4\zeta^2 L A^3+2\mu A^2\nonumber\\
&(\bar{s} L^2 \zeta^2+34)-2~\zeta^2 LA-{63\mu})+(H^2-1)A^2H^2(J+ML)(12L\zeta^2 A^5-24\mu A^4~(2+ \nonumber\\
&\bar{s} L^2\zeta^2)+22LA^3\zeta^2+4\mu A^2(35+2\bar{s} \zeta^2 L^2)-10AL\zeta^2-84\mu))],\\
\mathcal{J}=&L_{0}-P_{0} \nonumber\\
=&\dfrac{\pi}{2\mu L^2\zeta^2 A^{5}H^2(A^2+1)}[(H^2-1)A^2H^2(J+ML)(12L\zeta^2 A^5-24\mu A^4 (2+ \bar{s} L^2\zeta^2) \nonumber\\
& +22LA^3\zeta^2+4\mu A^2(35+2\bar{s} \zeta^2 L^2)-10AL\zeta^2-84\mu))-4h_{0}(A^4-1)(2H^2-1) \nonumber\\
&(LM-J)(-8\mu A^4+4\zeta^2 L A^3+2\mu A^2(\bar{s} L^2 \zeta^2+34)-2\zeta^2 LA-{63\mu})],
\end{align}
where $\mathcal{M}$ and $\mathcal{J}$ are the mass and angular momentum of WBTZ black holes.

\section{Entropy matching}\label{secc4}
The WBTZ black hole solution in ADM form is given as  \cite{Aggarwal:2020igb}, \cite{Kraus:2005zm}
\begin{equation}\label{metricADM}
ds^{2}=-N(r)^{2}dt^2+\dfrac{dr^2}{f(r)^2}+R(r)^2(N^{\phi}(r)dt+d\phi)^{2},
\end{equation}
with
\begin{align}
N^2(r)&=-\dfrac{4(2H^2-1)(J-ML)(16J^2L^2-8ML^2r^2+r^4)}{L(16H^2L^2J^2+H^2r^4-4Lr^2(ML+J(2H^2-1)))},\\
f^2(r)&=\dfrac{16J^2}{r^2}-8M+\dfrac{r^2}{L^2},\\
R^2(r)&=-\dfrac{16H^2J^2L^2+H^2r^4-4Lr^2(2H^2J-J+ML)}{4L(ML-J)},\\
N^{\phi}(r)&=\dfrac{H^2r^4-8MH^2L^2r^2-16JL^2(J(H^2-1)+LM(1-2H^2))}{L(H^2r^4+16H^2J^2L^2-4Lr^2(ML+J(2H^2-1)))}.
\end{align}
Taking $\xi=\partial_{t}+\frac{r_{-}}{Lr_{+}} \partial_{\phi}$ and given \eqref{metricADM}, the entropy of black hole in GMG is obtained as\cite{Wald1,Wald2}
\begin{small}
 \begin{align}\label{eqqsbulk}
 S^{GMG}=&\dfrac{\pi^{2}}{4\mu L^2\zeta^2 A^{5}H^2(A^2+1)\sqrt{M^2L^2-J^2}} [ 4h_{0}(A^4-1)(2H^2-1)(LM-J)(-8\mu A^4+\nonumber\\
& 4\zeta^2 L A^3+2\mu A^2(\bar{s} L^2 \zeta^2+34)-2\zeta^2 LA-{63\mu})+(H^2-1)A^2H^2(J+ML)(12L \nonumber\\
&\zeta^2 A^5- 24\mu A^4(2+\bar{s} L^2\zeta^2)+22LA^3\zeta^2+4\mu A^2(35+2\bar{s} \zeta^2 L^2)-10AL\zeta^2-84\mu)\nonumber\\
&\sqrt{ML^2+L\sqrt{M^2L^2-J^2}}+((H^2-1)A^2H^2(J+ML)(12~L\zeta^2 A^5-24\mu A^4(2+\nonumber\\
&\bar{s} L^2\zeta^2)+22LA^3\zeta^2+4\mu A^2(35+2\bar{s} \zeta^2 L^2)-10AL\zeta^2-84\mu))-4h_{0}(A^4-1)(2H^2 \nonumber\\
&-1)(LM-J)(-8\mu A^4+4\zeta^2 L A^3+2\mu A^2(\bar{s} L^2 \zeta^2+34)-2\zeta^2 LA-{63\mu}))\nonumber\\
&\sqrt{ML^2-L\sqrt{M^2L^2-J^2}} ],
 \end{align}
 \end{small}
where $r_{\pm}$ are the horizons of black holes (solutions of the equation $f(r) = 0$) and are given by
\begin{equation}
r_{\pm}=2\sqrt{GL}\sqrt{LM\pm\sqrt{L^2M^2-J^2}}.
\end{equation}
The Hawking temperature and angular velocity of the black hole are given as
\begin{equation}
T=\dfrac{r_{+}^{2}-r_{-}^{2}}{2\pi r_{+}L^2}=\dfrac{2\sqrt{L^2M^2-J^2}}{\pi L^{\frac{3}{2}}\sqrt{ML+\sqrt{M^2L^2-J^2}}},
\end{equation}
and
\begin{equation}\label{eqangu}
\Omega=\dfrac{r_{-}}{Lr_{+}}=\dfrac{\sqrt{ML-\sqrt{L^2M^2-J^2}}}{L\sqrt{ML+\sqrt{M^2L^2-J^2}}}.
\end{equation}
As expected, the above thermodynamic quantities (\ref{eqmassang}-\ref{eqangu}) satisfy the first law
\begin{equation}
d\mathcal{M}=TdS+\Omega d\mathcal{J}.
\end{equation}
Here, we want to obtain the entropy from the field theory side by counting the degeneracy of states in the boundary. The first step is to define the vacuum of the theory. The usual method to obtain the vacuum solution is to enhance the local symmetries to global symmetries which imposes $2\pi$ periodicity in $\phi$ (for more details see \cite{Aggarwal:2020igb}).
Therefore, a particular vacuum solution is obtained by setting $J=0, M=-1/8G$ as \cite{Aggarwal:2020igb}
\begin{equation}
ds_{vac}^{2}=(L^2+r^2)(2H^2r^2+L^2(2H^2-1))\dfrac{dt^2}{L^4}+\dfrac{L^2dr^2}{L^2+r^2}+4H^2r^2(L^2+r^2)\dfrac{dtd\phi}{L^3}+\left(r^2+\dfrac{2H^2r^4}{L^2}\right)d\phi^2.
\end{equation}
We see that in the case of $H=0$, the metric becomes global AdS$_{3}$. We have two values of GMG charges that the mass charge is 
\begin{align}
\mathcal{M}=&\dfrac {\pi}{8{L}^{2}\mu( {A}^{2}+1) {A}^{5}{\zeta}^{2}
{H}^{2}}(-16\mu\,{h_{0}}\,{A}^{8}-16\,L{\zeta}
^{2}{A}^{7}{H}^{2}{h_{0}}+12\,\mu\,{A}^{6}{\zeta}^{2}\bar{s}\,{L}^{
2}{H}^{4}-12\,\mu\,{A}^{6}{\zeta}^{2}\bar{s}\,{L}^{2}\nonumber\\
&{H}^{2}-8\,\mu
\,{A}^{6}{\zeta}^{2}\bar{s}\,{L}^{2}{H}^{2}{h_{0}}+4\,\mu\,{A}^{6}{
\zeta}^{2}\bar{s}\,{L}^{2}{h_{0}}-272\,\mu\,{A}^{6}{H}^{2}{h_{0}}+
11\,L{\zeta}^{2}{A}^{5}{H}^{2}-4\,L{\zeta}^{2}{A}^{5}{h_{0}} \nonumber\\
&-70
\,\mu\,{A}^{4}{H}^{4}+70\,\mu\,{A}^{4}{H}^{2}-110\,\mu\,{A}^{4}{h_{0}
}+5\,L{\zeta}^{2}{A}^{3}{H}^{4}-5\,L{\zeta}^{2}{A}^{3}{H}^{2}-8
\,L{\zeta}^{2}{A}^{3}{h_{0}}+42\,\nonumber\\
&{A}^{2}\mu\,{H}^{4}-42\,{A}^{2}
\mu\,{H}^{2}-136\,{A}^{2}\mu\,{h_{0}}-252\,\mu\,{h_{0}}\,{H}^{2}+8\,
L{\zeta}^{2}{A}^{5}{H}^{2}{h_{0}}-4\,\mu\,{A}^{4}{\zeta}^{2}
\bar{s}\,{L}^{2}{H}^{4}+\,\nonumber\\
&4\mu\,{A}^{4}{\zeta}^{2}\bar{s}\,{L}^{2}{H}
^{2}+220\,\mu\,{A}^{4}{H}^{2}{h_{0}}+16\,L{\zeta}^{2}{A}^{3}{H}^{2
}{h_{0}}+8\,{A}^{2}\mu\,{\zeta}^{2}\bar{s}\,{L}^{2}{H}^{2}{h_{0}}-
4\,{A}^{2}\mu\,{\zeta}^{2}\bar{s}\,{L}^{2}{h_{0}}\nonumber\\
&+272\,{A}^{2}\mu\,
{H}^{2}{h_{0}}-8\,L{\zeta}^{2}{h_{0}}\,A{H}^{2}+126\,\mu\,{h_{0}
}-6\,L{\zeta}^{2}{A}^{7}{H}^{4}+6\,L{\zeta}^{2}{A}^{7}{H}^{2}+8
\,L{\zeta}^{2}{A}^{7}{h_{0}}+ \nonumber\\
&24\,\mu\,{A}^{6}{H}^{4}+4\,L{\zeta
}^{2}{h_{0}}\,A+32\,\mu\,{h_{0}}\,{A}^{8}{H}^{2}-24\,\mu\,{A}^{6}{H}
^{2}+136\,\mu\,{A}^{6}{h_{0}}-11\,L{\zeta}^{2}{A}^{5}{H}^{4}
),
\end{align}
and the angular charge is
\begin{align}\label{eqang89vac}
\mathcal{J}=&-\dfrac {\pi}{8\mu( {A}^{2}+1) {H}^{2}{A}^{5}L{\zeta}^{2
}}(-16\mu{h_{0}}{A}^{8}-16L{\zeta
}^{2}{A}^{7}{H}^{2}{h_{0}}-12\,\mu\,{A}^{6}{\zeta}^{2}\bar{s}\,{L}^
{2}{H}^{4}+12\,\mu\,{A}^{6}{\zeta}^{2}\bar{s}\,{L}^{2}\nonumber\\
&{H}^{2}-8\,\mu
\,{A}^{6}{\zeta}^{2}\bar{s}\,{L}^{2}{H}^{2}{h_{0}}+4\,\mu\,{A}^{6}{
\zeta}^{2}\bar{s}\,{L}^{2}{h_{0}}-272\,\mu\,{A}^{6}{H}^{2}{h_{0}}-
11\,L{\zeta}^{2}{A}^{5}{H}^{2}-4\,L{\zeta}^{2}{A}^{5}{h_{0}}+ \nonumber\\
&70
\,\mu\,{A}^{4}{H}^{4}-70\,\mu\,{A}^{4}{H}^{2}-110\,\mu\,{A}^{4}{h_{0}
}-5\,L{\zeta}^{2}{A}^{3}{H}^{4}+5\,L{\zeta}^{2}{A}^{3}{H}^{2}-8
L{\zeta}^{2}{A}^{3}{h_{0}}-42{A}^{2}\mu \nonumber\\
&{H}^{4}+42{A}^{2}
\mu{H}^{2}-136{A}^{2}\mu\,{h_{0}}-252\mu{h_{0}}{H}^{2}+8
L{\zeta}^{2}{A}^{5}{H}^{2}{h_{0}}+4\,\mu\,{A}^{4}{\zeta}^{2}
\bar{s}\,{L}^{2}{H}^{4}-4\mu{A}^{4}{\zeta}^{2}\bar{s}\nonumber\\
&{L}^{2}{H}
^{2}+220\, \mu\,{A}^{4}{H}^{2}{h_{0}}+16\,L{\zeta}^{2}{A}^{3}{H}^{2
}{h_{0}}+8\,{A}^{2}\mu\,{\zeta}^{2}\bar{s}\,{L}^{2}{H}^{2}{h_{0}}-
4\,{A}^{2}\mu\,{\zeta}^{2}\bar{s}\,{L}^{2}{h_{0}}+272\,{A}^{2}\nonumber\\
&\mu\,
{H}^{2}{h_{0}}-8\,L{\zeta}^{2}{h_{0}}\,A{H}^{2}+126\,\mu\,{h_{0}
}+6\,L{\zeta}^{2}{A}^{7}{H}^{4}-6\,L{\zeta}^{2}{A}^{7}{H}^{2}+8
\,L{\zeta}^{2}{A}^{7}{h_{0}}-24\,\mu\,{A}^{6}{H}^{4}\nonumber\\
&+4\,L{\zeta
}^{2}{h_{0}}\,A+32\,\mu\,{h_{0}}\, {A}^{8}{H}^{2}+24\,\mu\,{A}^{6}{H}
^{2}+136\,\mu\,{A}^{6}{h_{0}}+11\,L{\zeta}^{2}{A}^{5}{H}^{4}
).
\end{align}
As can be seen from \eqref{eqang89vac}, the GMG angular momentum of vacuum solution does not equal to zero. This interesting result has been observed from other three dimensional gravitational theories containing parity-odd terms \cite{Carlip:1994hq},\cite{Setare:2021ref}.
In the quadratic ensemble, the warped Cardy formula takes the form
\begin{equation}\label{eeqq}
S_{WCFT}=4\pi\sqrt{-P_{0}^{vac}P_{0}}+4\pi\sqrt{-L_{0}^{vac}L_{0}},
\end{equation}
where the zero modes for vacuum metric become
\begin{align}
P_{0}^{vac}=&\dfrac {\pi h_{0}(-1+2H^{2})( A^2-1)( 8\mu A^{4}-4\zeta^{2
}L A^{3}-2A^{2}\mu\bar{s} L^{2}\zeta^{2}-68A^{2}\mu+
2\zeta^{2}LA+63\mu)}{4LH^{2}A^{5}\mu\zeta^{
2}},\\
L_{0}^{vac}=&-\dfrac {\pi( H^2-1)}{8L{A}^{3}\mu({A}^{2}+1) \zeta^{2}}(6{A}^{5}L\zeta^{2}-24\mu{A}^{4}-12{A}^{4}\mu \bar{s}{L}^
{2}\zeta^{2}+11\zeta^{2}L{A}^{3}+4{A}^{2}\mu\bar{s}{L
}^{2}\zeta^{2}+\nonumber\\
&70{A}^{2}\mu-5\zeta^{2}LA-42\mu
).
\end{align}
Inserting this in \eqref{eeqq}, one finds
\begin{align}
S_{WCFT}=&\dfrac{\sqrt{2}\pi^2}{H^2A^5\mu \zeta^2(A^2+1)}(2h_{0}(A^4-1)(2H^2-1)\sqrt{Y}(8\mu A^4-4\zeta^2 A^3-4\mu A^2(34+\bar{s}\zeta^2)+\nonumber\\
&4\zeta^2 A+63\mu)+H^2(H^2-1)A^2\sqrt{X}(6\zeta^2 A^5-12\mu A^4(2+\bar{s}\zeta^2)+11\zeta^2 A^3)+2\mu A^2 \nonumber\\
&(2\bar{s}\zeta^2+35)-5A\zeta^2-42\mu).
\end{align}
where $X=ML+J,Y=ML-J$.
After some manipulations, this expression matches the bulk thermodynamic WBTZ
entropy \eqref{eqqsbulk} provided that
\begin{equation}\label{eq94h0}
h_{0}=\dfrac{H^2A^2(H^2-1)(r_{+}+r_{-})(2X+\sqrt{2X}(r_{-}-r_{+}))C}{2(2H^2-1)(A^4-1)(r_{+}-r_{-}))D},
\end{equation}
where
\begin{align}
C&=6A^5\zeta^2-12\mu\bar{s} A^4\zeta^2-24\mu A^4+11\zeta^2 A^3+70\mu A^2+4\mu\bar{s} A^2\zeta^2-5\zeta^2 A-42\mu ,\\
D&=2Y+\sqrt{2Y}(r_{-}+r_{+})(8\mu{A}^{4}-4{\zeta}^{2}{A}^{3}-2\mu{A}^{2}\bar{s}{\zeta}^{2}-68\mu{A}^{2}+2{\zeta}^{2}A+63\mu .
\end{align}
In this section, we obtained the entropy of WBTZ via thermodynamical approach and the entropy of WCFT via Cardy formula. Finally, we showed that $S_{WCFT}=S_{WBTZ}$ if $h_{0}$ satisfied \eqref{eq94h0}.
\section{Conclusion}\label{secc5}
In this work, in the framework of general massive gravity, we studied the asymptotic symmetry algebra, the solution space, and the global charges using the ACDS boundary conditions in the quadratic ensemble. Under the mentioned boundary conditions, we construct the solution space with two arbitrary functions ($ f_{++}(x^{+}),h(x^{+})$) and two constants ($\varsigma_{+-},j_{++}$).
 The solution space is different from the counterpart for Einstein's gravity because the Cotton tensor and NMG part are not equal to zero and in the limit $\mu \to \infty$ and $\zeta \to \infty$, it gives the Einstein's solution space. 
Then, we obtained the asymptotic symmetry and their algebra ($\sigma$ generates an abelian algebra and $\epsilon$ generates the usual Witt algebra) by imposing FG gauge fixing on the metric and new boundary conditions. 
We obtained the integrable surface charges using the Iyer-Wald method by fixing a part of the solution space ($\delta j_{++}=\delta \varsigma_{+-}=0$).  We have obtained the centrally extended charge algebra which is $ Vir \otimes u(1)$ algebra 
with the central charges which are provided in \eqref{centralex}. When the TMG and NMG couplings tend to zero, the central charges tend to its Einstein counterpart.
We also show that the boundary counting of the degeneracy of states correctly reproduces the bulk thermodynamic entropy for WBTZ black holes. 
This confirms that the phase space has the same symmetries as that of a WCFT in the quadratic ensemble.

It would be interesting to extend the domain of validity of this new boundary conditions for the other 3D massive gravity theories (such as GMMG, EGMG) and different gauges (such as Bondi and Bondi-Weyl gauge). Also, it is interesting to compute the linearized energy excitations (energy of gravitons) in WAdS$_{3}$ at the chiral points of the theory. In addition, one can apply the mechanism of \cite{Ciambelli:2021nmv} to make the charges integrable. We leave these works for the future.

 \section*{Acknowledgements}
  We would like to thank the referee for his/her fruitful comments which help us to improve the presentation of the manuscript. SNS also would like to thank the School of Physics of the Institute for Research in Fundamental Sciences (IPM) for the research facilities.
\appendix

\section{WAdS$_{3}$ metric}
WAdS$_3$ black holes are different from the AdS$_{3}$ black holes and their properties are similar to the Kerr black holes. The asymptotic symmetry group of WAdS$_3$ black holes is the semi-direct product of a chiral Virasoro algebra with a $u(1)$ current. 
The metric of  WBTZ can be obtained by a deformation of the BTZ black hole spacetime as follows \cite{Banados:1992gq}-\cite{Israel:2004vv}
\begin{equation}\label{WBTZsol}
ds_{WBTZ}^{2}=ds_{BTZ}^{2}-2H^2\xi \otimes\xi,
\end{equation}
where
\begin{equation}
ds_{BTZ}^{2}=\dfrac{L^2r^2dr^2}{16J^2L^2-8ML^2r^2+r^4}+(4ML^2-r^2)dx^{+}dx^{-}+2L(LM+J)dx^{+2}+2L(LM-J)dx^{-2},
\end{equation}
with
\begin{equation}
\xi=-\dfrac{1}{\sqrt{2GL(LM-J)}}\partial_{-}.
\end{equation}
The metric of WBTZ is a solution of the GMG field equation if
\begin{small}
\begin{align}
H^2=&\dfrac{119\mu^2 -6\zeta^4 L^2 +2\zeta^2 L \sqrt{84\mu^2
+\zeta^2 L(9\zeta^2 -42\bar{s}\mu^2)}+14\bar{s}\mu^2\zeta^2 L^2}{294\mu^2},\\
\bar{\lambda} =&\dfrac{1}{3087\mu^4 L^4 \zeta^{2}}[2L\zeta^{2}(3L^2\zeta^{4}-14\bar{s} L^2\mu^2 \zeta^{2}-56\mu^{2})\sqrt{84\mu^2+\zeta^{2}L^2(9\zeta^{2}-42\bar{s})}-
18L^{4}\zeta^{8}\nonumber\\
&+126\bar{s} \mu^2 \zeta^{6}L^{4}+(252-147\bar{s}^2 \mu^2 L^2)\mu^2 L^2\zeta^{4}-4704\bar{s} \mu^4 \zeta^{2}L^2-2352\mu^4].
\end{align}
\end{small}
For $\bar{s}=(6\zeta^{2}L-17\mu)/(2\mu L^{2}\zeta^{2})$, $H$ becomes zero and $\bar{\lambda}=-(12\zeta^2L-35\mu)/4\mu\zeta^2L^4$ and the WBTZ metric becomes the BTZ metric.
For $\zeta \to \infty$, and assuming $L>0,\mu>0$ we have
\begin{equation}
\bar{\lambda}=\dfrac{-36\bar{s} +\bar{s}^2 \mu^2 L^2}{27L^2},\;\;\;\;H^2=\dfrac{1}{2}-\dfrac{\bar{s}^2\mu^2 L^2}{18},
\end{equation}
which are the same as \cite{Aggarwal:2020igb} for TMG. In the case of $\mu \to \infty$
\begin{equation}
\bar{\lambda}=-\dfrac{\bar{s}^2\zeta^4 L^4+32\bar{s}\zeta^2 L^2+16}{21\zeta^2 L^4},\;\;\;\;H^2=\dfrac{17}{42}+\dfrac{\bar{s} \zeta^2 L^2}{21},
\end{equation}
which are the same as \cite{NMGpaper} for NMG.
 
\end{document}